\documentclass[12pt,preprint]{aastex}
\usepackage{emulateapj5}
\usepackage{xspace}

\newcommand{\chandra}{{\it Chandra}\xspace} 
\newcommand{\xmm}{{\it XMM-Newton}\xspace}

\newcommand{\pulsar}{1E1207.4-5209\xspace}

\slugcomment{Accepted for publication in the Astrophysical Journal Letters}

\begin{document}

\title{Evidence for a Mid-Atomic-Number Atmosphere in the Neutron Star 1E1207.4-5209} 
\author{Charles J. Hailey and Kaya Mori}
\affil{Columbia Astrophysics Laboratory, New York, NY 10027}

\begin{abstract}

Recently  \citet{sanwal02} reported  the first  clear detection  of absorption
features  in an  isolated  neutron star,  \pulsar.  Remarkably their  spectral
modeling demonstrates that the  atmosphere cannot be Hydrogen. They speculated
that  the  neutron star  atmosphere  is indicative  of  ionized  Helium in  an
ultra-strong ($\sim1.5\times10^{14}$  G) magnetic  field. We have  applied our
recently developed atomic model \citep{mori02} for strongly-magnetized neutron
star atmospheres  to this problem.  We find that  this model, along  with some
simple atomic physics arguments,  severely constrains the possible composition
of  the atmosphere. In  particular we  find that  the absorption  features are
naturally  associated with  He-like  Oxygen or  Neon  in a  magnetic field  of
$\sim10^{12}$  G, comparable  to  the  magnetic field  derived  from the  spin
parameters of  the neutron  star. This interpretation  is consistent  with the
relative line strengths and widths  and is robust. Our model predicts possible
substructure in  the spectral  features, which has  now been reported  by \xmm
\citep{mereghetti02}. However  we show  the Mereghetti et  al. claim  that the
atmosphere is  Iron or  some comparable high-Z  element at $\sim10^{12}$  G is
easily ruled out by the \chandra and \xmm data.

\end{abstract}
\keywords{stars: neutron -- X-rays: stars -- stars: atmospheres -- individual:
1E1207.4-5209}

\section{Introduction}

A major goal of neutron star  (NS) research has remained unrealized despite 30
years of  effort - to determine  the fundamental properties  of the superdense
matter  in a  NS  interior, in  particular  its equation  of  state (EOS)  and
composition. There  are many approaches which  can be employed  in this effort
(\citet{vankerkwijk01}  and   references  therein),   but  one  of   the  most
intensively studied is to exploit the thermal radiation from isolated NS. This
radiation can  be used to deduce  information about the EOS  from neutron star
cooling theory \citep{tsuruta02,yakovlev99}.  Extracting information about the
NS interior is not straightforward,  however, since the observed spectrum does
not represent the  NS surface emission, but is  modified by radiative transfer
effects in the NS atmosphere.   The problem of unfolding the observed spectrum
and understanding NS interiors thus depends on deducing the composition of the
atmosphere.

NS atmospheres are interesting in their  own right. We know little about their
nature,  other   than  that  some   NS  probably  have   Hydrogen  atmospheres
\citep{pavlov02_2},  since they  fit  well to  the sophisticated  H-atmosphere
models that have been developed \citep{pavlov95_1, zavlin02}. The observations
have been silent on the question of whether non-Hydrogen atmospheres exist.
 
The  NS  \pulsar  was   discovered  by  \citet{helfand84}  with  the  Einstein
Observatory  and   is  associated  with   the  SNR  PKS   1209-51.  Subsequent
observations established it as a  prototypical radio silent NS associated with
a SNR \citep{bignami92, caraveo96}. ROSAT  and ASCA observations were fit with
a   black-body  spectrum  \citep{mereghetti96,   vasisht97}  and   a  Hydrogen
atmosphere    model   \citep{zavlin98_2}.     No    X-ray   pulsations    were
detected.    \chandra   detected   X-ray    pulsations   with    $P=0.424$   s
\citep{zavlin00}. While  this is perhaps  not surprising in light  of previous
work, the  small period derivative estimated  \citep{pavlov02_1} certainly was
unexpected.  The inferred surface B-field is $\sim(2-4)\times10^{12}$ G.

Even  more  remarkable,  \citet{sanwal02}  (hereafter  SZPT)  have  discovered
features at $\sim0.7$ keV and  $\sim1.4$ keV are required to obtain acceptable
fits  to the  spectrum (which  was fit  with a  $2.6\times 10^6$  K underlying
black-body continuum).  A feature of marginal significance  and unclear origin
was   also  noted   at  $\sim2$   keV.  A   subsequent  observation   by  \xmm
\citep{mereghetti02} has  provided some hint of substructure  in the features,
and has marginally detected the $\sim2$ keV feature seen by \chandra.

Three different phenomenological fits indicative of either absorption lines or
edges  were  used  by SZPT  to  fit  the  two  strong features.   Of  enormous
significance,  the  spectrum  was  shown  to  be  inconsistent  with  Hydrogen
atmosphere  models.  The intense  effort  which  has  gone into  the  Hydrogen
atmosphere  models  and their  success  in  explaining  other NS  observations
\citep{pavlov02_2}  lends  credence  to  the inevitable  conclusion  that  the
atmosphere  of \pulsar is  something other  than Hydrogen.   Unfortunately, as
pointed out  elsewhere \citep{zavlin02},  work on non-Hydrogen  atmospheres at
high B-fields is much less developed.  SZPT argued on various grounds that the
features  could  not  arise  via  cyclotron lines.  Instead  they  tentatively
suggested emission  from a  once ionized-Helium atmosphere  with a  B-field of
$1.5\times 10^{14}$  G.  This B-field  is inconsistent with that  derived from
the  spin parameters,  but SZPT  argue this  could be  due to  an off-centered
B-field or  glitches affecting the  $\dot{P}$ measurement. Others  have argued
for   a    cyclotron   line   solution   at    lower   B-field   \citep{xu02}.
\citet{mereghetti02} claimed an atmosphere of Iron or other high-Z elements at
a B-field  of $\sim10^{12}$ G, although  they did not actually  fit their \xmm
data to a model.

In  this   paper  we  offer   alternate  interpretations  of   these  spectral
features.  All  our  interpretations  involve atomic  transitions,  mainly  in
He-like  ions of  mid-Z  elements  at $B\sim10^{12}$  G,  consistent with  the
B-field  derived  from  the NS  spin  properties.  The  most likely  of  these
interpretations  is that  the  NS  atmosphere contains  Oxygen  or Neon;  most
noteworthy  is that {\it  all} our  models, whether  considering just  the two
strong features or including the third weak feature, demand mid-Z elements for
an acceptable solution.  Our model, combined with the  \xmm and \chandra data,
easily rule out the Iron and high-Z solutions of \citet{mereghetti02}.

Some comments are in order on  our approach. The atomic spectroscopy data used
in this analysis is based on  a novel approach for obtaining fast and accurate
solutions to  the Schr{\"o}dinger equation  for B-fields in the  Landau regime
(appropriate    for   all    cases   considered    here).     This   approach,
multiconfigurational,  perturbative,  hybrid,  Hartree,  Hartree-Fock  theory,
allows rapid  computation of transition energies and  oscillator strengths for
arbitrary atom,  ion, excitation state and  B-field (\citet{mori02}, hereafter
MH02a).  This  permits  a   complete  search  of  all  possible  spectroscopic
transitions consistent with the given line or edge energies.

While  it may  appear that  this  approach produces  an uninterestingly  large
number  of   potential  solutions,  we   demonstrate  in  a   companion  paper
(\citet{mori02_2}, hereafter MH02b)  that this is not the  case.  We show that
the  presence of  two or  more  line or  edge features  provides a  remarkable
robustness  to  a  host   of  poorly-understood  atomic  physics  effects  and
unambiguously restricts the atmosphere composition to mid-Z elements.

\section{\chandra data reduction}

\begin{figure}[h]
\epsscale{0.5}
\plotone{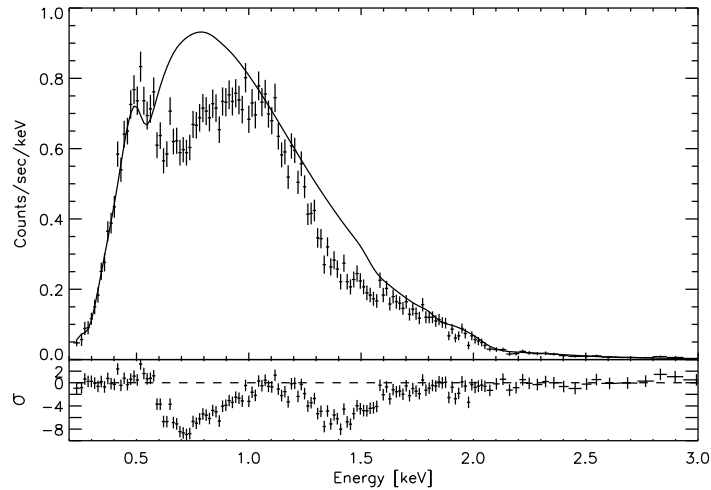}
\caption{\chandra/ACIS spectrum. Solid line is a blackbody model to clearly
illustrate the two absorption features. \label{fig_bb}}
\end{figure}

\begin{figure}[h]
\epsscale{1.0}
\plottwo{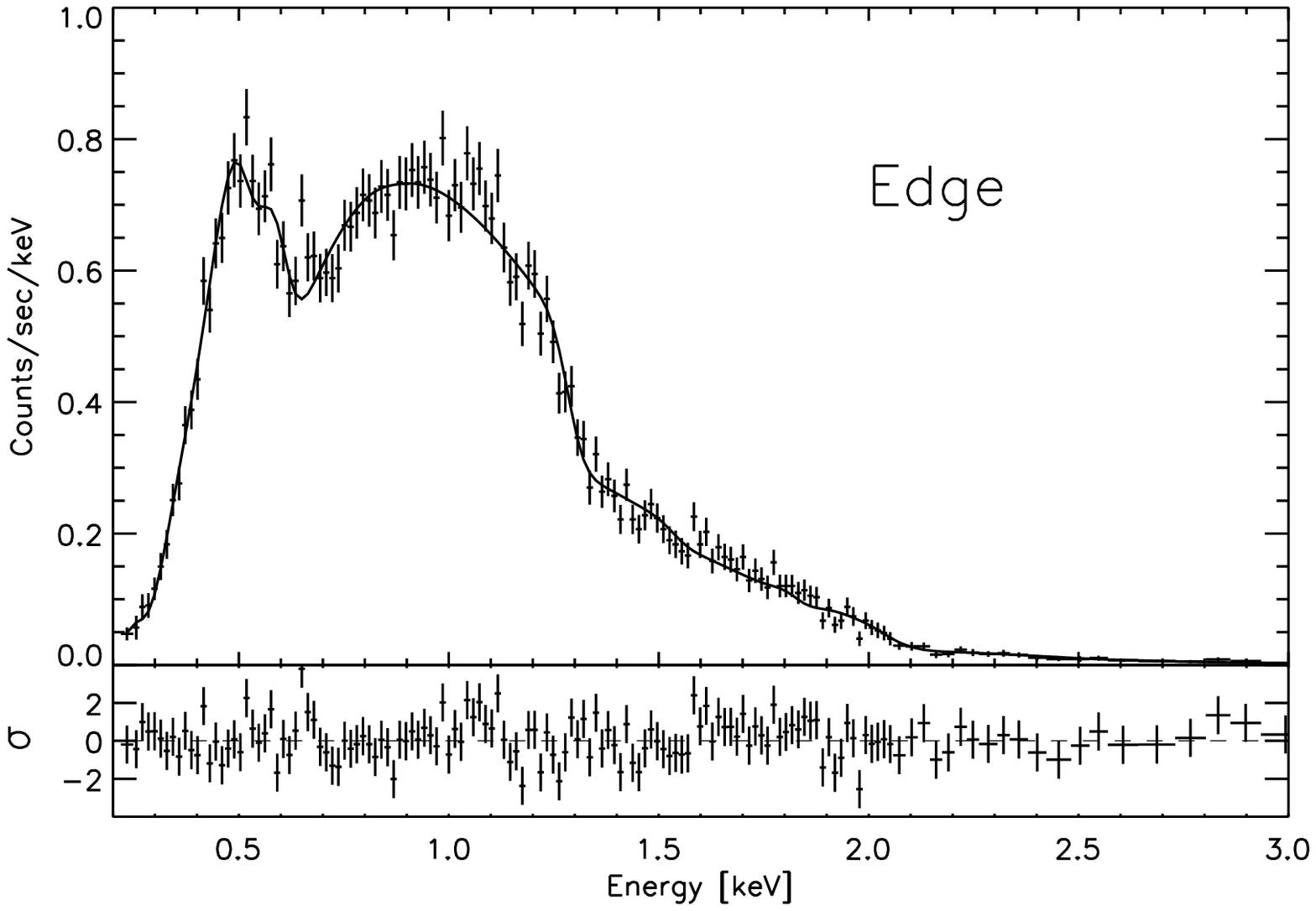}{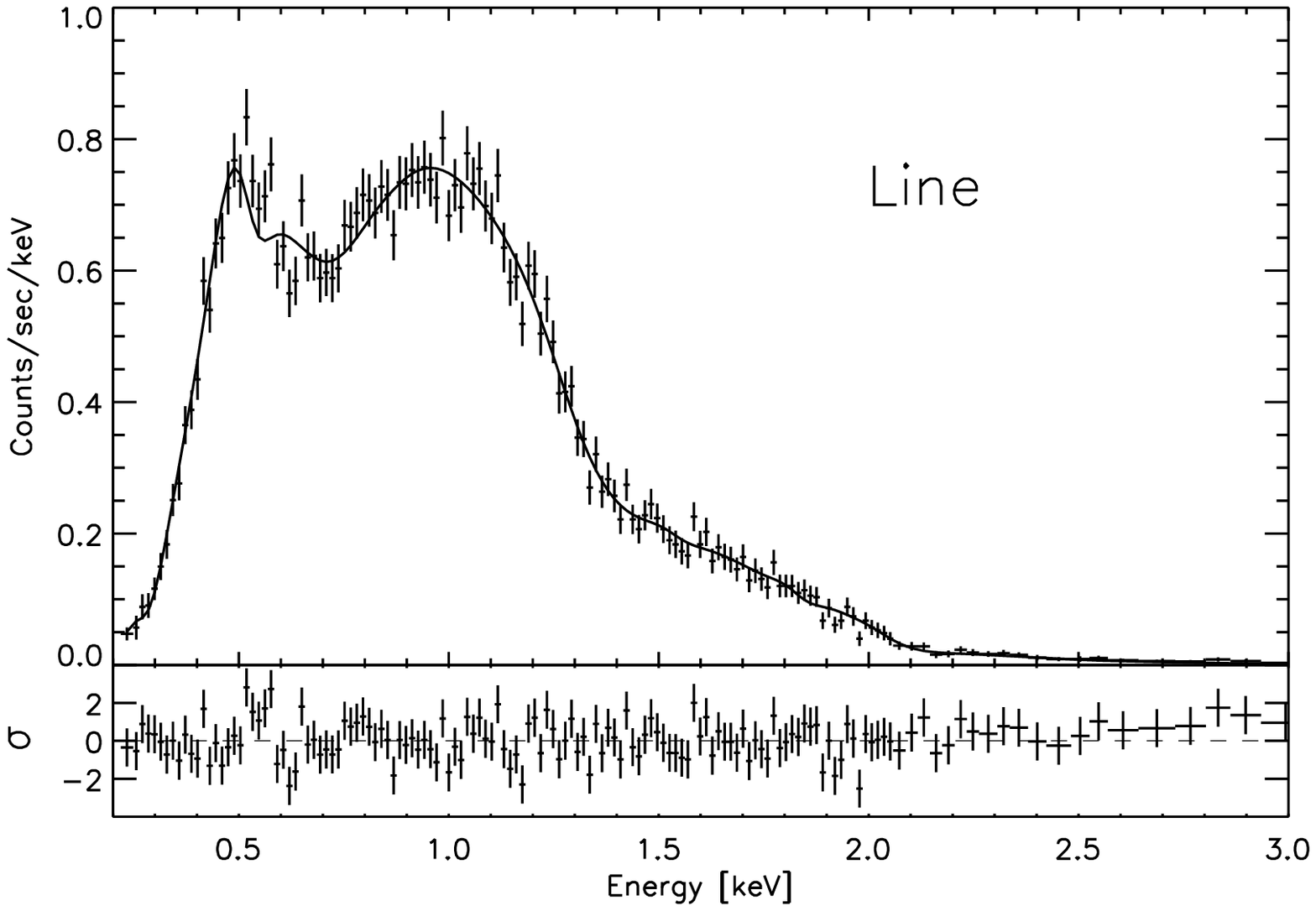} 
\caption{\chandra/ACIS spectrum fitted with the edge (left) and the gaussian
line model (right). For the edge model, we assumed the energy dependence of
the photo-absorption cross section to be $\sigma(E)\propto E^{-2}$ for the strong
magnetic field case \citep{miller92}. Other energy indices such as 3 yield
similar $\chi^2_\nu$ values. \label{fig_lineedge}}
\end{figure}

We  only briefly  mention our  data reduction  here, as  our  approach closely
follows that of  SZPT. Indeed, we emphasize that none  of our conclusions here
would be  modified if we simply  used the spectral line  parameters derived by
SZPT.   Our  analysis  is described  in  more  detail  in MH02b.  The  results
presented  here are for  phase-integrated spectra  only. Subsequent  work will
consider phase-resolved data, which can provide more information on the system
geometry. We fit the spectrum with two models: a blackbody with two absorption
edges (edge model) and a blackbody with two absorption lines (line model).  As
noted  by  SZPT,  these  spectral  features are  required  to  get  acceptable
fits. The fit  parameters for our models are given  in Table \ref{tab_fit}. We
show  the  raw  \chandra  spectrum  in figure  \ref{fig_bb}  overlaid  with  a
blackbody to clearly indicate the presence of the absorption features. Figures
\ref{fig_lineedge} show the results of our best fit line and edge models.

\section{General Spectroscopic Considerations}

We briefly  review the quantum numbers  of bound electrons in  B-fields in the
Landau  regime.  States are  labeled by  $(m,\nu)$ where  $m$ is  the magnetic
quantum  number and  $\nu$ is  the longitudinal  quantum number.   The $\nu=0$
states have larger binding energy than the $\nu>0$ states. Consistent with the
terminology of MH02a,  we call the former tightly-bound  states and the latter
loosely-bound states. Generally the ground  state of an ion is a tightly-bound
state.  In  the  X-ray  band  transitions  are  mostly  likely  to  be  either
photo-absorption (tightly-bound  state to continuum) or  absorption lines from
tightly-bound  to loosely-bound  state (tight-loose  transition).  Tight-tight
transitions appear  in the optical  band. Details of the  transition selection
rules are discussed elsewhere (MH02a,MH02b).

\section{Specific Atmosphere Solutions (Two Spectral Features)}

We first considered the two  strongest features at $\sim0.7$ keV and $\sim1.4$
keV and assume both features are due to atomic transitions in a single element
and ionization  state (Case  A). This effectively  defines a solution  for the
dominant absorbing element in terms of He-like ions. Li-like ions and those in
lower  ionization  states  are  not  dominant  absorbers  under  the  expected
atmospheric conditions  (MH02b). H-like ions cannot be  the dominant absorbers
because this is grossly incommensurate  with the features' relative widths and
strengths  even   under  conditions  far  removed   from  local  thermodynamic
equilibrium (MH02b). H-like  ions can, however, be present  since the relevant
spectral features  will be blended with  the He-like features.  In the He-like
ions, the electrons are making tight-loose transitions out of the $(0,0)(1,0)$
ground  state to  the  $(0,1)(1,1)$  excited states  respectively,  or to  the
continuum.  Neither tightly-bound states at higher magnetic quantum number nor
non-adjacent  tightly-bound states  are  acceptable. In  the  former case  the
feature energy  ratios cannot  be produced and  in the latter  case unobserved
features in the \chandra band would be produced.

We searched for  any combination of element, redshift  and B-field which could
give the  observed positions of the  spectral features. Our  B-field range was
$\sim10^{11}$ G  to $\sim2\times10^{14}$ G. In  the case of  edge solutions we
also permitted (small) pressure shifts  (details in MH02b). Line solutions are
insensitive  to pressure  shifts. The  properties of  the  solutions including
derived   values  of   the   B-field   and  redshift   are   shown  in   table
\ref{tab_summary}. The first two  solutions involve atomic line transitions in
He-like  Oxygen  and Neon  and  the  third  solution He-like  Neon  absorption
edges. A  similar Oxygen  solution was rejected  because its redshift  was too
low,  violating  causality  and  stability  conditions  for  the  NS  interior
\citep{lindblom84, haensel99}. For the Neon edge solution there is an electron
cyclotron line in the \chandra/\xmm energy band but it was made to overlap the
higher  energy  spectral  feature   by  introducing  a  small  pressure  shift
($\sim50-90$ eV). Such pressure shifts slightly affect the energy ratio of the
two edge features and serve to further constrain other elements and conditions
of the atmosphere  (MH02b).  The line and edge widths  are all consistent with
expectations  for   Stark  broadening  and   non-constant  B-field  broadening
respectively (MH02b).
   
Our case B  assumes that one spectral feature is an  atomic transition and the
second  is a cyclotron  line. A  1.4 keV  or 0.7  keV electron  cyclotron line
defines a  B-field of  $1.2\times10^{11}(1+z)$ G or  $0.6\times10^{11}(1+z)$ G
respectively. In the  former case solutions can only be  found for H-like ions
transitioning out of the $(0,0)$ ground  state to the $(0,1)$ excited state or
the continuum.  These are all mid-Z  solutions albeit with  lower B-field than
those considered above.  In the case of a 0.7 keV electron cyclotron line, the
B-field  is low enough  that our  perturbation method  breaks down  for higher
Landau  levels and  is  therefore  less accurate;  we  estimate that  putative
solutions  would involve  the elements  Beryllium or  Boron and  are therefore
implausible. If one of the features is a nuclear cyclotron line then the other
feature  must be  a  transition from  the  ground state  of  either Helium  or
Hydrogen. However  there is no combination  of $B$ and $z$  that can reproduce
the line positions. Thus nuclear cyclotron lines are ruled out.

Although Case B is formally allowed  we consider it highly unlikely. There are
difficulties in  attempting to reconcile the  depth and width  of the spectral
features  (MH02b). Nevertheless  it  is interesting  that  even this  scenario
including cyclotron lines only admits solutions involving mid-Z elements.

\section{Specific Atmosphere Models (Three Spectral Features)}

Table  \ref{tab_summary}  also shows  proposed  solutions  if  the feature  at
$\sim2$ keV is confirmed (Case C solutions). Again it is of importance to note
that only mid-Z solutions are acceptable.  We had to relax one of our previous
assumptions by assuming additional significant absorption from H-like ions. An
admixture  of  H-like  ions in  a  given  atmosphere  is quite  plausible  for
reasonable conditions of temperature and density (MH02b).

The first  case considered is identical  to the He-like  Oxygen/Neon case presented
above but  with the  addition of a  H-like Oxygen/Neon  absorption edge due  to the
$(0,0)$ ground state  to continuum transition. In this  model the other H-like
lines overlap the He-like lines so no additional features are produced. 
 
The second solution invokes photoabsorption  edges in He-like Neon for the two
lower energy  features and  the higher  energy feature is  due to  an electron
cyclotron line.

\section{Discussion}

Our most important conclusion is  that {\it all} viable solutions for B-fields
comparable  to  that  inferred  from  the NS  spin  parameters  require  mid-Z
atmospheres in  both the  two and three  spectral feature case.  Our solutions
which most  closely correspond to  the inferred B-field demand  He-like Oxygen
and Neon.

In contrast  to this paper,  SZPT proposed ultrahigh B-field  solutions. Their
Helium  atmosphere appears  problematic, however.  In order  to get  the right
feature positions  SZPT's H-like  Helium features must  of necessity  arise in
transitions from electrons in the $(0,0)$ or $(1,0)$ quantum states to $(0,1)$
and $(1,1)$ states.  The $(0,1)$ state is heavily depopulated  at the types of
ion densities one expects and the $(1,1)$ state is actually autoionized due to
the  large nuclear  cyclotron energy.  In fact  the motional  Stark  shift and
pressure effects in  Helium will likely destroy all  the loosely-bound states.
Consequently one  is forced to assume  the features are  absorption edges, and
this  leads  to  severe  difficulty  explaining the  relative  edge  strengths
(MH02b).

\citet{mereghetti02} did  not fit their  \xmm data, but  proposed nevertheless
that the  spectral features could be  attributed to Iron or  some other high-Z
element  at  a  field  $\sim10^{12}$  G.  Our current  work  rules  out  their
proposal. There is no combination  of B-field, redshift, element or ionization
states, even  with unphysical, arbitrary  pressure shifts, that  is consistent
with the \xmm and \chandra data.

The interpretation we present here makes an important prediction: the observed
features may show substantial  substructure when observed with higher spectral
resolution.   Transition  energies from  the  tightly  bound  ground state  to
loosely bound  final excited states  differing in longitudinal  quantum number
are  rather close  to each  other, and  may be  blended. Details  of predicted
energies and  transition assignments are presented in  MH02b. Whether detailed
atmospheric models  would lead to a  washing out of  this substructure through
broadening is  not clear, but its  detection would certainly  confirm that the
features are atomic transitions since we would not expect similar substructure
in cyclotron  lines. In fact, \citet{mereghetti02} claim  a marginal detection
of such substructure.

Concerning the  B-field derived from the  free parameter fits to  the data, we
 note that our Neon  and Oxygen line solutions are all within  a factor of two
 or three of the B-field derived from spin parameters. The discrepancy may lie
 in the  NS B-field geometry or  line blending. Line blending  will affect our
 B-field   (and  redshift)  estimate,   but  will   not  affect   our  element
 identifications.

The analysis here, along with that of  SZPT, begs the question of why there is
anything in the  atmosphere other than Hydrogen.  Oxygen  or Neon are expected
constituents  near the  NS mass  cut in  many models  of  supernova explosions
\citep{arnett96}. Indeed due to turbulent mixing in such explosions Oxygen can
be found  throughout the  core \citep{herant94}.  But  given that a  very thin
layer of Hydrogen  accreted from the interstellar medium  is enough to produce
an optically  thick Hydrogen atmosphere, we  are left to  ponder mechanisms by
which  such  accretion can  be  inhibited,  or  at least  permit  simultaneous
emission from  a mid-Z element. One  possibility is that the  timescale for an
optically thick  layer to accrete  onto the surface  is shorter than  that for
gravitational stratification in  the atmosphere \citep{vankerkwijk01}. In that
case,  even trace  amounts of  mid-Z elements  could lead  to  strong emission
\citep{pavlov02_2}. Optical  observations of  the associated SNR  indicate the
presence of  Oxygen and it  has been argued  that the unusually  high Galactic
latitude  of this  NS implies  the Oxygen  originated in  the  progenitor star
\citep{ruiz83}. Thus  a rich  source of Oxygen  is available.  Alternately the
mid-Z  atmosphere may  be associated  with the  NS mass  cut, and  a propeller
effect may be severely inhibiting  accretion of Hydrogen. We note that several
of our proposed solutions allow a Hydrogen-rich environment (MH02b).

 For  more progress to  be made  on \pulsar  it will  be necessary  to develop
complete models  of NS  atmosphere for mid-Z  elements incorporating  the more
sophisticated atomic data bases we have utilized here. This source has much to
tell us in the years  to come, especially when higher resolution spectroscopic
observations become available.

\acknowledgments{The authors thank David Helfand for helpful discussions and
a careful reading of the manuscript.}

\begin{deluxetable}{lrrr}
\tablewidth{0pt}
\tablecaption{Results from spectral fitting by the three models. \label{tab_fit}}
\tablehead{
\colhead{ } & \colhead{Edge} & \colhead{Gaussian} & \colhead{Lorentzian}}
\startdata 
$kT^{\infty}$ [eV] &$269\pm2$&$254\pm2$&$254\pm2$\\
$N_H$ [10$^{20}$cm$^{-2}$] & $4.5\pm0.3$ & $4.6\pm0.3$ & $5.1\pm0.4$\\
$R^{\infty}$ [km] \tablenotemark{a} & $1.68\pm0.04$&$1.80\pm0.05$&$1.91\pm0.07$\\
$E_1$ [eV]	&$626\pm5$	&$737\pm6$	&$737\pm6$\\
$\tau_1$\tablenotemark{b} &$0.60\pm0.05$&$0.46\pm0.03$&$0.57\pm0.03$\\
EW$_1$ [eV] & $338\pm28$ &$119\pm10$ &$327\pm27$\\
$E_2$ [eV] & $1298\pm10$ &  $1424\pm10$	& $1430\pm10$\\
$\tau_2$\tablenotemark{b} & $0.54\pm0.05$ &$0.47\pm0.04$ &$0.55\pm0.05$\\
EW$_2$ [eV]& $637\pm53$ &  $115\pm10$ &  $199\pm17$ \\
line (edge) ratio &$2.07\pm0.03$ & $1.93\pm0.03$&$1.94\pm0.03$\\
$\chi^2_\nu$\tablenotemark{c} & 1.07 & 1.01 &  1.10 \\
\enddata
\tablenotetext{a}{We assumed the distance $\sim$ 2.1 kpc for apparent radius.}
\tablenotetext{b}{Optical depth of edges and lines.} 
\tablenotetext{c}{142 degrees of freedom.}
\end{deluxetable}

\begin{deluxetable}{cccccc}
\tablewidth{0pt}
\tablecaption{Candidates for elements, magnetic field
strengths and gravitational redshifts. \label{tab_summary}}
\tablehead{\colhead{Case} & \colhead{Ion} & \colhead{Features} &
\colhead{$B_{12}$} & \colhead{Redshift} & \colhead{Pressure shift?}}
\startdata
A	&\ion{O}{7}& two lines & 0.55--0.75 & 0.06--0.21 	& No	\\
A	&\ion{Ne}{9}& two lines & 0.80--1.10    & 0.62--0.86 & No	\\
A	&\ion{Ne}{9}& two edges & 0.18--0.20    & 0.56--0.57 & Yes   \\ 
B	& \ion{C}{6}	& one edge \& cyclotron line &	0.21--0.22&0.70-0.73 & No\\
B	& \ion{N}{7}	& one line \& cyclotron line	& 0.13--0.14	& 0.10--0.13	& No	\\
B	& \ion{O}{8}	& one line \& cyclotron line	& 0.18-0.19	&
0.45-0.49	& No 	\\
C	& \ion{O}{7}, \ion{O}{8}&two lines \& one edge & 0.55--0.75 &
0.06--0.21& No \\ 
C	& \ion{Ne}{9}, \ion{Ne}{10}& two lines \& one edge & 0.80--1.10 &
0.62--0.86 & No \\
C	& \ion{Ne}{9}	& two edges \& cyclotron line&	0.26--0.28& 0.59--0.60 	& Yes	\\
\enddata
\end{deluxetable}

\end{document}